\documentclass[12pt]{article}
\begin{document}
\def\bfone{\relax{\rm 1\kern-.35em 1}}\let\shat=\hat
\def\hat{\widehat}

%
\def\cS{{\cal K}}
\def\IE{\relax{{\rm I\kern-.18em E}}}
\def\cE{{\cal E}}
\def\rt{{\cR^{(3)}}}
\def\IGam{\relax{{\rm I}\kern-.18em \Gamma}}
\def\IGa{\IA}
\def\cV{{\cal V}}
\def\Rt{{\cal R}^{(3)}}
\def\tft#1{\langle\langle\,#1\,\rangle\rangle}
\def\IA{\relax{\hbox{{\rm A}\kern-.82em {\rm A}}}}
\def\hata{{\shat\a}}
\def\hatb{{\shat\b}}
\def\hatA{{\shat A}}
\def\hatB{{\shat B}}
\def\bv{{\bf V}}
\def\Fb{\overline{F}}
\def\nablab{\overline{\nabla}}
\def\Ub{\overline{U}}
\def\Db{\overline{D}}
\def\zb{\overline{z}}
\def\eb{\overline{e}}
\def\fb{\overline{f}}
\def\tb{\overline{t}}
\def\Xb{\overline{X}}
\def\Vb{\overline{V}}
\def\Cb{\overline{C}}
\def\Sb{\overline{S}}
\def\delb{\overline{\del}}
\def\Gammab{\overline{\Gamma}}
\def\Ab{\overline{A}}
\def\Anh{A^{\rm nh}}
\def\alphab{\bar{\alpha}}
\def\cy{Calabi--Yau}
\def\cabg{C_{\alpha\beta\gamma}}
\def\B{\Sigma}
\def\Bh{\hat \Sigma}
\def\Kh{\hat{K}}
\def\Knh{{\cal K}}
\def\A{\Lambda}
\def\Ah{\hat \Lambda}
\def\R{\hat{R}}
\def\V{{V}}
\def\T{T}
\def\Gammah{\hat{\Gamma}}
\def\twot{$(2,2)$}
\def\K{K\"ahler}
\def\rat{({\theta_2 \over \theta_1})}
\def\lv{{\bf \omega}}
\def\w{w}
\def\CP{C\!P}
\def\o#1#2{{{#1}\over{#2}}}
\def\eq#1{(\ref{#1})}

\def\ib{{\bar \imath}}
\def\jb{{\bar \jmath}}
\def\Im{{\rm Im ~}}
\def\Re{{\rm Re ~}}
\def\IP{\relax{\rm I\kern-.18em P}}
\def\arccosh{{\rm arccosh ~}}
%
\font\cmss=cmss10 \font\cmsss=cmss10 at 7pt
\def\twomat#1#2#3#4{\left(\matrix{#1 & #2 \cr #3 & #4}\right)}
\def\inbar{\vrule height1.5ex width.4pt depth0pt}
\def\IC{\relax\,\hbox{$\inbar\kern-.3em{\rm C}$}}
\def\IG{\relax\,\hbox{$\inbar\kern-.3em{\rm G}$}}
\def\IB{\relax{\rm I\kern-.18em B}}
\def\ID{\relax{\rm I\kern-.18em D}}
\def\IL{\relax{\rm I\kern-.18em L}}
\def\IF{\relax{\rm I\kern-.18em F}}
\def\IH{\relax{\rm I\kern-.18em H}}
\def\II{\relax{\rm I\kern-.17em I}}
\def\IN{\relax{\rm I\kern-.18em N}}
\def\IP{\relax{\rm I\kern-.18em P}}
\def\IQ{\relax\,\hbox{$\inbar\kern-.3em{\rm Q}$}}
\def\bfzero{\relax\,\hbox{$\inbar\kern-.3em{\rm 0}$}}
\def\IR{\relax{\rm I\kern-.18em R}}
\def\ZZ{\relax\ifmmode\mathchoice
{\hbox{\cmss Z\kern-.4em Z}}{\hbox{\cmss Z\kern-.4em Z}}
{\lower.9pt\hbox{\cmsss Z\kern-.4em Z}}
{\lower1.2pt\hbox{\cmsss Z\kern-.4em Z}}\else{\cmss Z\kern-.4em
Z}\fi}
\def\IU{\relax\,\hbox{$\inbar\kern-.3em{\rm U}$}}
\def\bfone{\relax{\rm 1\kern-.35em 1}}
\def\dop{{\rm d}\hskip -1pt}
\def\real{{\rm Re}\hskip 1pt}
\def\trace{{\rm Tr}\hskip 1pt}
\def\ii{{\rm i}}
\def\diag{{\rm diag}}
\def\sch#1#2{\{#1;#2\}}

\newcommand{\ft}[2]{{\textstyle\frac{#1}{#2}}}
\newcommand{\QED}{{\hspace*{\fill}\rule{2mm}{2mm}\linebreak}}
\def\dop{{\rm d}\hskip -1pt}
\def\bfone{\relax{\rm 1\kern-.35em 1}}
\def\bfzero{\relax{\rm I\kern-.18em 0}}
\def\inbar{\vrule height1.5ex width.4pt depth0pt}
\def\IC{\relax\,\hbox{$\inbar\kern-.3em{\rm C}$}}
\def\ID{\relax{\rm I\kern-.18em D}}
\def\IF{\relax{\rm I\kern-.18em F}}
\def\IK{\relax{\rm I\kern-.18em K}}
\def\IH{\relax{\rm I\kern-.18em H}}
\def\II{\relax{\rm I\kern-.17em I}}
\def\IN{\relax{\rm I\kern-.18em N}}
\def\IP{\relax{\rm I\kern-.18em P}}
\def\IQ{\relax\,\hbox{$\inbar\kern-.3em{\rm Q}$}}
\def\IR{\relax{\rm I\kern-.18em R}}
\def\IG{\relax\,\hbox{$\inbar\kern-.3em{\rm G}$}}
\font\cmss=cmss10 \font\cmsss=cmss10 at 7pt
\def\ZZ{\relax\ifmmode\mathchoice
{\hbox{\cmss Z\kern-.4em Z}}{\hbox{\cmss Z\kern-.4em Z}}
{\lower.9pt\hbox{\cmsss Z\kern-.4em Z}}
{\lower1.2pt\hbox{\cmsss Z\kern-.4em Z}}\else{\cmss Z\kern-.4em
Z}\fi}
\def\i{\rm i}
\def\a{\alpha} \def\b{\beta} \def\d{\delta}
\def\e{\epsilon} \def\c{\gamma}
\def\G{\Gamma} \def\l{\lambda}
\def\L{\Lambda} \def\s{\sigma}
\def\cA{{\cal A}} \def\cB{{\cal B}}
\def\cC{{\cal C}} \def\cD{{\cal D}}
\def\cF{{\cal F}} \def\cG{{\cal G}}
\def\cH{{\cal H}} \def\cI{{\cal I}}
\def\cJ{{\cal J}} \def\cK{{\cal K}}
\def\cL{{\cal L}} \def\cM{{\cal M}}
\def\cN{{\cal N}} \def\cO{{\cal O}}
\def\cP{{\cal P}} \def\cQ{{\cal Q}}
\def\cR{{\cal R}} \def\cV{{\cal V}}\def\cW{{\cal W}}
%
%
%
\def\crr{\crcr\noalign{\vskip {8.3333pt}}}
\def\tilde{\widetilde}
\def\bar{\overline}
\def\us#1{\underline{#1}}
\let\shat=\hat
\def\hat{\widehat}
\def\hyp{\vrule height 2.3pt width 2.5pt depth -1.5pt}
\def\square{\mbox{.08}{.08}}

\def\Coeff#1#2{{#1\over #2}}
\def\Coe#1.#2.{{#1\over #2}}
\def\coeff#1#2{\relax{\textstyle {#1 \over #2}}\displaystyle}
\def\coe#1.#2.{\relax{\textstyle {#1 \over #2}}\displaystyle}
\def\half{{1 \over 2}}
\def\shalf{\relax{\textstyle {1 \over 2}}\displaystyle}
\def\dag#1{#1\!\!\!/\,\,\,}
\def\to{\rightarrow}
\def\notin{\hbox{{$\in$}\kern-.51em\hbox{/}}}
\def\shdot{\!\cdot\!}
\def\ket#1{\,\big|\,#1\,\big>\,}
\def\bra#1{\,\big<\,#1\,\big|\,}
\def\equaltop#1{\mathrel{\mathop=^{#1}}}
\def\Trbel#1{\mathop{{\rm Tr}}_{#1}}
\def\inserteq#1{\noalign{\vskip-.2truecm\hbox{#1\hfil}
\vskip-.2cm}}
\def\attac#1{\Bigl\vert
{\phantom{X}\atop{{\rm\scriptstyle #1}}\phantom{X}}}
\def\exx#1{e^{{\displaystyle #1}}}
\def\del{\partial}
\def\delbar{\bar\partial}
\def\nex#1{$N\!=\!#1$}
\def\dex#1{$d\!=\!#1$}
\def\cex#1{$c\!=\!#1$}
\def\eg{{\it e.g.}} \def\ie{{\it i.e.}}
%
\def\cS{{\cal K}}
\def\IE{\relax{{\rm I\kern-.18em E}}}
\def\cE{{\cal E}}
\def\rt{{\cR^{(3)}}}
\def\IGam{\relax{{\rm I}\kern-.18em \Gamma}}
\def\IGa{\IA}
\def\cV{{\cal V}}
\def\Rt{{\cal R}^{(3)}}
\def\tft#1{\langle\langle\,#1\,\rangle\rangle}
\def\IA{\relax{\hbox{{\rm A}\kern-.82em {\rm A}}}}
\def\hata{{\shat\a}}
\def\hatb{{\shat\b}}
\def\hatA{{\shat A}}
\def\hatB{{\shat B}}
\def\bv{{\bf V}}
\def\Fb{\overline{F}}
\def\nablab{\overline{\nabla}}
\def\Ub{\overline{U}}
\def\Db{\overline{D}}
\def\zb{\overline{z}}
\def\eb{\overline{e}}
\def\fb{\overline{f}}
\def\tb{\overline{t}}
\def\Xb{\overline{X}}
\def\Vb{\overline{V}}
\def\Cb{\overline{C}}
\def\Sb{\overline{S}}
\def\delb{\overline{\del}}
\def\Gammab{\overline{\Gamma}}
\def\Ab{\overline{A}}
\def\Anh{A^{\rm nh}}
\def\alphab{\bar{\alpha}}
\def\cy{Calabi--Yau}
\def\cabg{C_{\alpha\beta\gamma}}
\def\B{\Sigma}
\def\Bh{\hat \Sigma}
\def\Kh{\hat{K}}
\def\Knh{{\cal K}}
\def\A{\Lambda}
\def\Ah{\hat \Lambda}
\def\R{\hat{R}}
\def\V{{V}}
\def\T{T}
\def\Gammah{\hat{\Gamma}}
\def\twot{$(2,2)$}
\def\K{K\"ahler}
\def\rat{({\theta_2 \over \theta_1})}
\def\lv{{\bf \omega}}
\def\w{w}
\def\CP{C\!P}
\def\o#1#2{{{#1}\over{#2}}}
\def\eq#1{(\ref{#1})}
\newcommand{\be}{\begin{equation}}
\newcommand{\ee}{\end{equation}}
\newcommand{\ba}{\begin{eqnarray}}
\newcommand{\ea}{\end{eqnarray}}
\newtheorem{definizione}{Definition}[section]
\newcommand{\bd}{\begin{definizione}}
\newcommand{\ed}{\end{definizione}}
\newtheorem{teorema}{Theorem}[section]
\newcommand{\bth}{\begin{teorema}}
\newcommand{\eth}{\end{teorema}}
\newtheorem{lemma}{Lemma}[section]
\newcommand{\blem}{\begin{lemma}}
\newcommand{\elem}{\end{lemma}}
\newcommand{\brr}{\begin{array}}
\newcommand{\err}{\end{array}}
\newcommand{\nn}{\nonumber}
\newtheorem{corollario}{Corollary}[section]
\newcommand{\bcorol}{\begin{corollario}}
\newcommand{\ecorol}{\end{corollario}}
\def\twomat#1#2#3#4{\left(\begin{array}{cc}
 {#1}&{#2}\\ {#3}&{#4}\\
\end{array}
\right)}
\def\twovec#1#2{\left(\begin{array}{c}
{#1}\\ {#2}\\
\end{array}
\right)}
\thispagestyle{empty}
\begin{titlepage}
\thispagestyle{empty}
\begin{flushright}
CERN-TH/98-388\\
KUL-TF-98/51\\
December 1998\\
\end{flushright}
\vskip 2.cm
\begin{center}
{\Large\bf On short and long $SU(2,2/4)$ multiplets
 in the $AdS/CFT$ correspondence
\footnote{Work
 supported in part by EEC under TMR contract ERBFMRX-CT96-0045 (LNF Frascati
 and K.U.Leuven)
 and by DOE grant DE-FG03-91ER40662}}
\vskip 2.cm
{ Laura Andrianopoli }\\
{\it Institute for Theoretical Physics, KULeuven, Celestijnenlaan 200 D,\\
 B-3001 Leuven, Belgium}\\
{Sergio Ferrara}\\
{\it  CERN Theoretical Division, CH 1211 Geneva 23, Switzerland.}
\end{center}

\begin{abstract}
We analyze short and long multiplets which appear in the OPE expansion of
``chiral'' primary operators in $N=4$ Super Yang--Mills theory.
Among them, higher spin long and new short multiplets appear,
having the interpretation, in the AdS/CFT correspondence,
 of string states and supergravity multiparticle states respectively.

We also analyze the decomposition of long multiplets under $N=1$ supersymmetry,
as a possible tool to explore other supersymmetric deformations of IIB string
on $AdS_5 \times S_5$.
\end{abstract}
\end{titlepage}
\section{Introduction}
The recent advances in putting forward the correspondence
between superstring theories on $AdS_{d+1}$ backgrounds and
$d$-dimensional superconformal field theories \cite{m} deeply rely
on the supergravity approximation of string theory which, on
 the field theory side, roughly corresponds to retain a certain subclass
of conformal operators which have ``finite'' conformal dimension
in the limit of large t'Hooft parameter $g^2 N$ (for the case of 3--branes
with $AdS_5$ geometry).
\par
On the other hand, if one would like to explore further this connection
at finite $N$, then stringy corrections to the supergravity approximation
must be taken into account, as for example the inclusion of $R^4$ terms or
 D-instanton mediated processes \cite{bg}.

On the side of the superconformal sector, stringy corrections may correspond
to the appearence, in the OPE algebra, of conformal primary operators with
$g^2N$ dependent, i.e. not quantized, dimension \cite{m,fz}.

It is in fact a known fact that these operators do indeed occur, in
perturbative supersymmetric Yang--Mills theory \cite{afgj}, in the OPE of the stress
tensor multiplet, superconformal symmetry then requiring the extension of this
analysis to the entire class of operators related by superconformal
transformations \cite{a}.

Another context in which such multiplets play a role is the analysis of
deformation of a given superconformal field theory \cite{ls,kw}
 or, on the supergravity
side, choosing the vacuum of the corresponding theory on $AdS_5$ 
\cite{gppz,dz,kpw}.

In this analysis, the notion of conformal dimension of a given operator is a
property of the supergravity vacuum, since in the $AdS/CFT$ correspondence
the conformal dimension is mapped into the AdS energy $E_0$, the latter
depending on the particular extremum of the AdS supergravity potential.

However in the conformal field theory framework it is possible to explore
vacuum solutions beyond the $AdS_5$ supergravity analysis because it is
possible to add conformal deformations corresponding to massive K--K states
or even string states which are not present in the supergravity potential which
only involves ``massless scalars'', i.e. those scalars which belong to the
$n=8$ ``massless'' supergravity multiplet in the $AdS$ bulk.

It is therefore relevant to classify sequences of multiplets and their
corresponding scalar content in the wider context of $n=4$ superconformal
field theory, which allows one to consider general classes of operators
other than the K--K tower.
In this context $n=4$ superconformal symmetry plays an important role because
it allows to separate short and long multiplets in a rather simple way.

The former correspond to the K--K excitations of type II string theory on
$AdS_5 \times S_5$, the latter should correspond to string states, since they
cannot have a supergravity interpretation.
It should be emphasised that such separation makes sense for $n=8$ supergravity
but not for $n=2$.
Indeed in the latter case it is possible to have K--K states with anomalous
dimensions, i.e. long multiplets in the pure supergravity context \cite{cfn}.
This is for instance what generally happens for the K--K recurrences of
theories on backgrounds of the type $AdS_5 \times X$ where $X$ is a manifold
preseving $n=2$ supersymmetry \cite{kw}.
In these cases the only necessarily short multiplets are the hypermultiplets
but not the graviton, gravitino or vectors K--K recurrences.

\section{A class of long supermultiplets and their scalar content}
The classification of UIR reps. of highest weights can be done, in certain
cases, with the oscillator method developed by G\"unaydin et al. \cite{gs}.

This construction is especially powerful to classify the so-called ``chiral
primary'' $n=4$ operators in the $AdS_5/CFT_4$ correspondence.

The interpretation of the K--K spectrum of type IIB on $AdS_5 \times S_5$ in
terms of UIR of the $SU(2,2/4)$ superalgebra with the oscillator construction
was obtained by G\"unaydin and Marcus \cite{gm}.
The correspondence of these ``short multiplets'' to a particular class of
superfield operators of the $n=4$ superconformal field theory, introduced by
Howe and West \cite{hw}, was elucidated in ref. \cite{af}.
These short superfields correspond to massless ($p=2$) and massive ($p>2$)
reps. of the $SU(2,2/4)$
superalgebra with a lowest component scalar in the $(0,p,0)$ rep. of $SU(4)$
and with a total number $\frac{1}{12} p^2(p^2-1)2^8$ of states, with highest
spin 2 in the $(0,p-2,0)$ $SU(4)$ rep.

The shortening corresponds to the fact that only half of the 16 $\theta$'s
in the superfield expansion produces states.
These short multiplets only exist for ``quantized'' dimensions i.e. for lowest
component scalars with energy $E_0 =p$.

On the other hand long multiplets, in which all $\theta$'s components produce
states although one can obtain them by multiplying short multiplets, also
exist for values of the energy which are not quantized, but rather an arbitrary
real number, only subject to a certain unitarity bound ($E_0 \ge 2+J_1+J_2$
for the lowest component of the class of long supermultiplets we are going
to consider) \cite{ff,fgg}.

These multiplets have multiplicity proportional to $2^{16}$, the
 proportionality factor being related to a finite dimensional representation
of the maximal compact subgroup $SO(4) \times SO(2) \times SU(4)$ of the
bosonic
subalgebra $SO(4,2) \times SU(4)$ in the $SU(2,2/4)$ superalgebra.

The simplest of these long multiplets, discussed in ref. \cite{fz}, is the real
scalar superfield, with maximum spin 4 in a $SU(4)$ singlet.

This multiplet is contained in the tensor product of ``two'' singleton
supermultiplets as the first component is a scalar state
\begin{equation}
s=Tr (\phi_\ell \phi^\ell )
\end{equation}
where $\phi_\ell$ is the $\theta =0$ component of the Yang--Mills ``singleton''
superfield \cite{hst}.

This is the $n=4$ version of the so-called Konishi-multiplet.

In the free field theory limit, i.e. when the singleton superfield is abelian,
this multiplet becomes short and in fact can be obtained as the product of two
conjugate singleton reps. with maximum spin $(2,0)$ and $(0,2)$ respectively.

These are the singleton supermultiplets described in ref. \cite{gmz}, with 
lowest
spin $(J_L-1,0),(0,J_R-1)$, each with multiplicity
$(2J_L+1)2^4$.
Multiplying two conjugated singletons $(J_L-1,0) \times (0,J_R-1)$ one obtains
the massless reps. with minimum spin $(J_L-1, J_R-1)$ and maximum spin
$(J_L+1, J_R+1)$ and energy, for the lowest component, $E_0 =J_L+J_R$.
The number of states is $2^8 (2(J_L+J_R)+1)$.
All $(J_1,J_2)$ representations with $J_1 \cdot J_2 \neq 0$ inside these
multiplets correspond to ``conserved operators'' on the boundary \cite{fz}.

The free field Konishi-multiplet corresponds to $J_L=J_R=1$ and indeed contains
a singlet scalar only, with $E_0=J_L+J_R=2$.

Note that all massless higher spin supermultiplets of interest, given in the
table 12 of ref. \cite{gmz}, are obtained by taking $J_L=J_R >1$, and none of them
contains scalar fields.

Let us now consider long multiplets. These are obtained by multiplying two
singleton interacting multiplets, i.e. two singleton non-abelian Yang--Mills
multiplets.

In this case one gets $2^{16}$ states for the ``massive'' Konishi-multiplet.
All $(J_L,J_R)$ reps. with $J_L\cdot J_R \neq 0$ occurring in this multiplet are no
longer conserved, as it was in the free field case.
The analysis of this multiplet in component notation was given in ref.
\cite{fz}.

It is useful to report here the components with their $SU(4)$ assignment and
 the $AdS$ energy, i.e. the conformal weight.
The analysis for each field in the supermultiplet is described in the tables
 \ref{0,0}-\ref{2,3/2}.
It is immediate to see, from table \ref{0,0}, that this multiplet contains 
many scalars with
different $SU(4)$ and conformal weight assignment.

The important point is that the spectrum of this multiplet is independent from
the value of $E_0$ of its lowest component, which can then be lifted to any value
$E_0 \geq 2$.
We see that there are new scalar states with $E_0+1, E_0+2, \cdots , E_0+8$.
All these scalars are analogous of $F$ and $D$ terms which would otherwise
vanish in the free field theory.

Let us now consider higher spin superfields of the type appearing in the OPE
of the stress tensor multiplet in $n=4$ SYM theory \cite{a}.
These superfields are believed to be superfields whose highest spin component
is a $(J+1,J+1)$ $SU(4)$ singlet (non conserved) operator  whose dimension, in
 the free field theory limit, would be $E_0 = 2(2+J)$.

It is obvious that this is the massive generalization of the massless
supermultiplet obtained by multiplying two conjugate singletons with lowest
spin component $(J-1, 0)$, $(0,J-1)$ and integral $J>1$.
These multiplets are simply obtained by tensoring the scalar supermultiplet
(the massive spin 4 superfield) with a rep. $(J-1,J-1)$ of $SL(2,C)$.

The even spin $J>4$ are then obtained by tensoring with $(1,1)$, $(2,2)$ etc.

Since the highest spin in the scalar superfield is $(2,2)$, one will obtain
scalars only up to a superfield which transforms as a $(2,2)$ of $SL(2,C)$.

The scalars of the spin 6 and spin 8 superfields are all ``irrelevant'' from a
conformal point of view in the sense that their naive dimension is
$\ell \geq 6$.
They all vanish in the free field theory limit, where these multiplets become
``massless''.

The superfield which contains relevant and marginal operators is the
Konishi-multiplet, other than the stress tensor multiplet and the $p=3,4$
 massive short multiplets.
There are other superfields that contain scalars with naive dimension
 corresponding to relevant or marginal deformations.
These are the long multiplets contained in the lowest reps. of the symmetric
tensor product of $p$ singletons with $p\leq 4$.
For $p=2$ this is exactly the Konishi superfield since
\begin{equation}
(6\times 6)_S=20_R+1
\end{equation}
For $p=3$ we have:
\begin{equation}
(6\times 6\times 6)_S=50 + 6
\end{equation}
so there is a long multiplet whose lowest component has naive dimension
$E_0=3$ in the $6$ of $SU(4)$.
It has a scalar partner with $E_0=4$ in the $15+15+45 + \bar{45}$.

For $p=4$ we have
\begin{equation}
(6\times 6\times 6\times 6)_S=105+20_R+1
\end{equation}
so we have two long multiplets with lowest component $E_0=4$ in the $20_R$ and
 $1$ of $SU(4)$.

All these multiplets, having as lowest component a scalar field, have the same
structure of the Konishi-multiplet (max spin 4) where all states have $SU(4)$
reps. tensored with the rep. of the lowest component, i.e. the $6$ for $p=3$
and the $20+1$ for $p=4$.

Note that the long multiplets of the type of Konishi (max spin 4) but maked
up with more than two $\phi_\ell$'s in the free field theory limit do not
correspond to massless higher spin fields, but rather to massive ones.
This is because $E_0=2+J_L+J_R$ is not satisfied for these multiplets
\cite{ff,fgg}.
This of course also implies that these multiplets will have the same structure
irrespectively wheter the theory is abelian or not.
From the $AdS$ point of view of UIR reps. of $O(4,2)$, this has to do with
the fact that the product of more than two singletons gives rise to massve
representations.

In the $SU(N)$ Yang--Mills theory these multiplets with higher power of
 $\phi$'s are 
also distinguished by simple traces or multiple traces in the YM gauge group.
We will return to this in section 4.
\subsection{Spectrum of scalars in $J_{max}=4,6,8$ multiplets}
These are the multiplets which contain, as maximum spin, the
spin $4,6,8$ $SU(4)$ singlets.
The scalar spectrum is \footnote{We denote by $\ell^0$ the ``naive'' dimension
of these operators, since in general they have anomalous dimension}:
\begin{itemize}
\item{
$J_{max}=4$ ($\ell_4$ denotes the dimension of its lowest weight component):
\begin{tabular}{c|c}
$SU(4)$ & $E_0$,  ($\ell^0_4 =2$)  \\
\hline
1 & $\ell_4$, $\ell_4$ +2, $3(\ell_4+4)$, $\ell_4$+6, $\ell_4$+8 \\
10 & $\ell_4$+1, $2(\ell_4+3)$, $2(\ell_4+5)$, $\ell_4+7$ \\
$\bar{10}$ & $\ell_4$+1, $2(\ell_4+3)$, $2(\ell_4+5)$, $\ell_4+7$ \\
$20_R$ & $2(\ell_4+2)$, $3(\ell_4+4)$, $2(\ell_4+6)$ \\
15 & $\ell_4+2$, $\ell_4+4$, $\ell_4+6$ \\
84 & $\ell_4+2$, $\ell_4+4$, $\ell_4+6$ \\
\end{tabular}
}
\item{
$J_{max}=6$:
\\
\begin{tabular}{c|c}
$SU(4)$ &  $E_0$,  ($\ell^0_6 =4$)  \\
\hline
1& $\ell_6+2$,$3(\ell_6+4)$, $\ell_6+6$ \\
15 & $\ell_6+2$, $4(\ell_6+4)$, $\ell_6+6$ \\
$20_R$ & $\ell_6+2$,$3(\ell_6+4)$, $\ell_6+6$ \\
6 & $2(\ell_6+3)$, $2(\ell_6+5)$\\
10 & $2(\ell_6+3)$, $2(\ell_6+5)$\\
$\bar{10}$ & $2(\ell_6+3)$, $2(\ell_6+5)$\\
64 & $2(\ell_6+3)$, $2(\ell_6+5)$\\
45 & $\ell_6+4$ \\
$\bar{45}$ & $\ell_6+4$\\
84 & $\ell_6+4$ \\
\end{tabular}
}
\item{
$J_{max}=8$:
\\
\begin{tabular}{c|c}
$SU(4)$ &  $E_0$,  ($\ell^0_8 =6$)  \\
\hline
1& $\ell_8+4$ \\
\end{tabular}
}
\end{itemize}
The only multiplet containing scalars with $\ell^0 \leq 4$ is the Konishi
multiplet.

The total spectrum of scalar reps. is from $E_0 =2$ to $E_0=10$.

For $J_{max} >8$ no scalars exist.

Short multiplets

\begin{tabular}{c|l}
 & $SU(4)$ \\
\hline
Super-Konishi deformations:& \\
$E_0=2$ & 1 \\
$E_0=3$ & 10 ($\bar{10}$)\\
$E_0=4$ & 1, $2\times 20_R$, 15, 84 \\
\hline
$p=2$ Supercurrent deformations: & \\
$E_0=2$ & $20_R$ \\
$E_0=3$ & 10 ($\bar{10}$)\\
$E_0=4$ & $1+ \bar{1}$ \\
\hline
$p=3$ Supercurrent deformations: &  \\
$E_0=3$ & 50\\
$E_0=4$ & 45 ($\bar{45}$) \\
\hline
$p=4$ Supercurrent deformations: & \\
$E_0=4$ & 105 \\
\end{tabular}
\section{$n=1$ multiplet analysis for scalar operators with 
naive dimension $\ell = 3 , 4$}
The analysis of scalar operators with $E_0 \leq 4$
\footnote{We consider here the naive canonical dimension}
 in $n=4$ Yang--Mills theory
can be obtained in a rather straightforward way by knowing the relevant,
massless
and massive reps. of the $SU(2,2/4)$ algebra.

We first remind the result in the analysis of the short K--K multiplets:
These multiplets are classified by a quantum number $p$, such that the lowest
component scalar has $E_0 =p$ and is in the $(0,p,0)$ of $SU(4)$.

The analysis therefore includes the 3 multiplets with $p=2,3,4$.
The first multiplet is the supergravity multiplet and therefore its 42
scalars are:
\begin{equation}
42 = 20_R \, (E_0=2), 10 + \bar{10} \,(E_0=3), 1+\bar{1} \, (E_0=4)
\end{equation}
These are the scalars which appear in the gauged supergravity potential.
It is understood, according to the previous analysis, that the values of $E_0$
in the above formula refer to the $SU(4) \times AdS_5$ invariant vacuum in
 which $<20_R> = <10> =0$.

There are two extra scalar operators in the $p=3$ sector:
\begin{equation}
50 \, (E_0=3), 45 \, (E_0=4)
\end{equation}
and finally $105 \, (E_0=4)$ in the $p=4$ sector.

Note that all scalars, except the $SU(4)$ singlet, necessarily break
$n=4$ supersymmetry.

The scalars which preserve $n=1$ supersymmetry can be found by decomposing
$SU(4) \to SU(3) \times U(1)$ and looking at scalars which are the highest
component
of $n=1$ superfields, according to the classification of ref. \cite{fz}.
For the relevant representations we have:
\begin{eqnarray}
20_R &\to & 6(\frac{4}{3}) + \bar{6} (-\frac{4}{3}) + 8(0) \nonumber \\
10 & \to & 1(2) + 3(\frac{2}{3}) + 6(-\frac{2}{3}) \nonumber \\
50 &\to & 15 (\frac{2}{3}) + \bar{15} (-\frac{2}{3}) +
 10 (2) +\bar{10} (-2) \nonumber \\
45 &\to & 3(\frac{8}{3}) + \bar{3}(\frac{4}{3}) + 6(\frac{4}{3}) +
  8(0) + 10(0) + 15(-\frac{4}{3})
\end{eqnarray}
The analysis of these operators was performed in ref. \cite{fz}.
They contain, in particular, the chiral superfields corresponding to a
symmetric superpotential $10(2)$, a singlet superpotential $1(2)$,
a supersymmetric mass term $6(\frac{4}{3})$.

Let us now consider the long multiplets.
The only interesting multiplets are the ones up to spin 4, since
higher spin supermultiplets have scalars with too high dimension.

We consider the basic Konishi superfield with $\ell^0=2,3,4$, corresponding to
the traces in the $Tr(\phi_{\ell_1} \times \phi_{\ell_2} \cdots )$
product up to fourth order polynomial.
This multiplet has scalars with:
\begin{eqnarray}
E_0=\ell && SU(4) \mbox{ singlet}\nonumber\\
E_0=\ell +1 && 10 + \bar{10} \nonumber\\
E_0=\ell + 2 && 1 + 2\times 20_R + 15 + 84
\end{eqnarray}
For $\ell^0 =2$ we have both relevant and marginal.\\
For $\ell^0 =4$ we have only one marginal deformation in the $20_R + 1$.\\
For the scalar superfield (with $\ell^0=2$) there is a supersymmetric deformation
in the $10$, $\bar{10}$, corresponding to a superpotential which reduces to a 
$SU(3)$ singlet $f_{\alpha\beta\gamma}\phi^\alpha_\ell \phi^\beta_m 
\phi^\gamma_n$.
The higher dimensional operators correspond to the following objects:
\begin{equation}
f_{\alpha\beta\gamma}\psi^\alpha_A \psi^\beta_B \phi^\gamma_{[CD]} \, ,
\quad f_{\alpha\beta\gamma}\phi^\beta_\ell \phi^\gamma_m
f_{\alpha\epsilon\delta}\phi^\epsilon_p \phi^\delta_q
\end{equation}
which give rise precisely to the rep. content listed above.

There is another long multiplet which starts with scalars in the $6$
 ($\ell^0=3$),
but this does not give chiral $n=1$ multiplets among its components.

Finally, there is also a massive $J_{max}=5$ multiplet with scalars with $\ell
=4$,
but this multiplet does not appear in the OPE of the stress tensor,
because of symmetry reasons.
\subsection{$n=1$ massive representations}
The long multiplets of the $SU(2,2/1)$ algebra can be obtained from the lowest dimensional 
representation, by tensoring with a $(J_1,J_2)$ $SL(2,C)$ representation.

Since $n=4$ long multiplets have anomalous dimension, in their decomposition 
to $n=1$ they cannot contain ``short'' $n=1$ ``chiral'' multiplets.

Therefore the content of a generic massive $n=1$ $AdS_5$ multiplet is:
\begin{eqnarray}
&&{\cal D}(E_0,J_1,J_2,q), {\cal D}(E_0+\half,J_1+\half,J_2,q+1),\nonumber\\
&& 
{\cal D}(E_0+\half,J_1-\half,J_2,q+1),{\cal D}(E_0+\half,J_1,J_2+\half ,q-1),\nonumber\\
&& 
{\cal D}(E_0+\half,J_1,J_2-\half , q-1), {\cal D}(E_0+1,J_1,J_2 ,q+2),
\nonumber\\
&&
{\cal D}(E_0+1,J_1,J_2 ,q-2),
{\cal D}(E_0+1,J_1+\half,J_2+\half ,q),\nonumber\\
&& 
{\cal D}(E_0+1,J_1+\half,J_2-\half ,q),
{\cal D}(E_0+1,J_1-\half,J_2+\half ,q),\nonumber\\
&&
{\cal D}(E_0+1,J_1-\half,J_2-\half ,q),
{\cal D}(E_0+\frac{3}{2},J_1+\half,J_2,q+1),\nonumber\\
&&
{\cal D}(E_0+\frac{3}{2},J_1-\half ,J_2 ,q-1),
{\cal D}(E_0+\frac{3}{2},J_1,J_2 +\half ,q+1),\nonumber\\
&&
{\cal D}(E_0+\frac{3}{2},J_1,J_2-\half ,q-1),{\cal D}(E_0 +2,J_1,J_2,q)
\end{eqnarray}
This multiplet is described, in the AdS/CFT correspondence, by a ``superfield''
(of conformal dimension $\ell=E_0$):
\begin{equation}
V^{q,E_0}_{\alpha_1 , \cdots , \alpha_{2J_1},
\dot\alpha_1 , \cdots , \dot\alpha_{2J_2}}(x,\theta,\bar\theta)
\end{equation}
with $2J_1, 2J_2$ symmetrized $SL(2,C)$ indices.
\\
For massless representations, which occur when $E_0=\ell = 2+J_1+J_2$,
then $V$ is ``conserved'' ($J_1\cdot J_2 \neq 0$):
\begin{equation}
D^{\alpha_1\dot\alpha_1}V^{q,E_0}_{\alpha_1 , \cdots , \alpha_{2J_1},
\dot\alpha_1 , \cdots , \dot\alpha_{2J_2}}=0.
\end{equation}
When $J_1\cdot J_2 = 0$, the multiplet can obey another ``shortening 
condition'', i.e. the chirality constraint:
\begin{equation}
\bar D^{\dot\alpha_1}V^{q,E_0}_{\alpha_1 , \cdots , \alpha_{2J}}(x,\theta , 
\bar\theta )=0.
\end{equation}
(This demands $\ell =q$ where $q$ is the $U(1)$ R-charge).
For $\ell =q=1+J$ the chiral multiplet describes a singleton representation
in $AdS_5$.

For $\ell=q>1+J$ one gets both massless and massive conformal ``chiral''
excitations in the bulk.

\subsection{$n=1$ decomposition of long $n=4$ multiplets}
In order to analyze the $N=1$ content of $n=4$ super Yang--Mills theories it is
important to analyze the UIR's of the $SU(2,2/4)$ algebra in terms of $n=1$
superconformal representations.

For a generic $n=4$ long multiplet, this amounts to decompose it in terms
of $n=1$ representations as discussed in the previous section.

In this section we will report such a decomposition for the $n=4$
 Konishi-multiplet, which corresponds to the smallest $n=4$ long multiplet and
 is of physical interest since it appears in the OPE of $n=4$ Yang--Mills
 theory.

On general grounds, this multiplet has a maximum spin content $J_1=J_2=2$, with
$\ell_{J_1=J_2=2}=\ell +4$, where $\ell$ is the conformal dimension of the
superfield (i.e. the dimension of its lowest $\theta$ component).

This immediately implies that there will be a highest $n=1$ massive rep.
 of the type $V_{J_1=J_2=\frac{3}{2}}(x,\theta , \bar\theta )$, i.e. a spin 3
$n=1$ massive superfield.
Then the $N=4$ Konishi-multiplet will decompose in a hierarchy of $n=1$ 
superfields $\sum
 V_{J_1,J_2}^{R_{J_1J_2}}$, with $J_1 \leq \frac{3}{2} ,J_2 \leq \frac{3}{2}$, 
where $R_{J_1J_2}$ will be the suitable $SU(3)\times U(1)$ representation
which appears in the decomposition of ``highest weight'' $SU(4)$
 representations.

 Recently different supergravity vacua on $AdS_5$ have been interpreted as
possible conformal deformations of $n=4$ Yang--Mills theory \cite{kw,gppz,dz}.

We will only refer on those $n=1$ multiplets containing scalars with naive
 dimension $\ell \leq 4$.
The only $n=1$ massive multiplets containing scalars are those for which the
lowest component is $(0,0),(\half , 0), (0,\half ), (\half ,\half )$, so
only these types of multiplets will be analyzed.

The lowest dimensional state of the $n=4$ Konishi-multiplet is a real scalar
with $E_0 =\ell$. It then follows that the $n=1$ multiplet with the lowest value
of $E_0$ will be a $V_{J_1=J_2=0}$ multiplet with $E_0=\ell$.

The next lowest $E_0$ multiplet will then be a
 $V^{R_{\half , 0}}_{J_1=\half , J_2 =0}$
multiplet with $E_0=\ell +\frac{1}{2}$ and $R_{\half , 0}=  3(-\frac{1}{3})$, 
and so on.
In this way we get a unique decomposition of the $n=4$ long multiplets in terms
 of $n=1$ ones, that is we have, for the $n=1$ multiplets with lowest
 component up to $E_0=\ell +2$:
\\

\begin{tabular}{c|c}
\hline
$E_0$ & $n=1$ multiplet \\
\hline
$\ell $  & $V^{1(0)}_{(0,0)}$\\
$\ell +\half $  & $V^{3(-\frac{1}{3})}_{(\half,0)}$;
 $V^{\bar 3(\frac{1}{3})}_{(0,\half )}$   \\
$\ell +1 $  & $V^{6(-\frac{2}{3})}_{(0,0)}$;$V^{\bar 6(\frac{2}{3})}_{(0,0)}$;
$V^{1(0)}_{(\half ,\half )}$; $V^{8(0)}_{(\half ,\half )}$ \\
$\ell +\frac{3}{2} $  & $V^{\bar 3(\frac{1}{3})}_{(\half,0)}$;
$V^{\bar{15}(\frac{1}{3})}_{(\half,0)}$;
$V^{8(-1)}_{(\half,0)}$;
 $V^{ 3(-\frac{1}{3})}_{(0,\half )}$;
$V^{ 15(-\frac{1}{3})}_{(0,\half )}$;
$V^{8(1)}_{(0,\half )}$    \\
$\ell +2 $  & $V^{\bar 6(-\frac{4}{3})}_{(0,0)}$;
$V^{ 6(\frac{4}{3})}_{(0,0)}$;
$V^{1(0)}_{(0,0)}$; $V^{8(0)}_{(0,0)}$;
$V^{27(0)}_{(0,0)}$ \\
\end{tabular}

\section{Comments on the $n=4$ OPE Expansion}
General properties of superconformal covariant OPE expansions have been
investigated in several papers \cite{afgj,a,hw,o}.

From the general results on the $n=4$ case one can draw some conclusions.
If one denotes by $O_{SG}$, $O_{KK}$, $O_{ST}$, operators in the $n=4$
super Yang-Mills theory that correspond to, respectively, supergravity
(i.e. $AdS$-massless), K--K, and string states, then the 3-point functions of
the type $\langle O_{SG}O_{SG}O_{ST}\rangle$, $\langle O_{KK}O_{KK}O_{ST}
\rangle$ are nonvanishing. In ref. \cite{a} an analog of the class of $O_{ST}$
 operators
contributing to the OPE of the stress tensor was given.

By $n=4$ superconformal symmetry, this analysis can be further generalized by
stating the following result:

The OPE expansion of the $n=4$ $O_{SG}$ multiplet contains the $O_{SG}$
multiplet itself, together with all long multiplets whose maximum spin is an
$(s,s)$ representation of $SL(2,C)$ with $s\geq 1$ 
\cite{a}.

The lowest-energy (dimension) component of this $AdS$ massive representation
is $E_0=\delta +2(s-1)$, with spin $(s-2,s-2)$.
 For $\delta =0$ this
representation has a massless limit, for which the $(s,s)$ tensors
(with $AdS$ energy $E_0=2(s+1)$) are conserved.
This is the case of the OPE expansion in the free field theory.

The Konishi-multiplet just corresponds to the $s=2$ case.

Notice that in the free field theory, sequences of such representations exist
for integral $\delta$, corresponding to multilinear (rather than bilinear)
composites in the Yang-Mills superfield.  These multilinear operators, from
an $AdS$ point of view, correspond to massive (rather than massless) $AdS$
representations, obtained by tensoring more than two singleton representations.
Let us call such sequences of higher-$\delta$ level $O^\delta_{ST}$ ($\delta=0$
corresponds to the Konishi-multiplet). These sequences are expected to
appear naturally in $\langle O_{KK}O_{KK}O^\delta_{ST}\rangle$, as it is
implied by the free field theory result.
In particular, all K--K operators, corresponding to the same $p$-level, will
contain the $\delta=0$ multiplet $O_{ST}$, which is just the Konishi
superfield, in their OPE, i.e. $\langle O^p_{KK}O^p_{KK}O_{ST}\rangle\neq 0$.

Note that in the Yang--Mills theory one can get other sequences of multilinear
operators by taking not single traces.

For example $(Tr(\phi_\ell \phi_m ) -traces)(Tr(\phi_p \phi_q ) -traces)
	     \to 105 + 84 + 20 +1$
would give rise both to short (105) and long (84 + 20 + 1) multiplets which
are  neither	K--K nor string states.

These states should correspond to multiparticle states of pure supergravity
\cite{o}.

The existence of short multiplets not corresponding to K--K
states can be seen directly by working in harmonic superspace \cite{gikos}.
In this case the K--K states are (G-)analytic (F-)holomorphic fields $Tr W^p$
\cite{hw}.

Now it is obvious that if we take, at level $p$, $\Pi_{i=1}^r Tr(W^{q_i})$, 
($\sum_r q_r =p$) this superfield is also G-analytic, F-holomorphic, i.e.
a short ``chiral primary'' in the $n=4$ superconformal language.
In the $n=1$ formalism, this is related to the fact that chiral primary
superfields form a closed algebraic ring \cite{o}.

The non-vanishing superfields for which $\{q_r\}\neq p$ will be called 
multiparticle states.

For a $SU(N)$ gauge theory, single trace operators exist up to level $p=N$.
Therefore if $p>N$ we would be in a situation in which only multiparticle 
states would exist.

Since in supergravity theory K--K states exist for arbitrary $p$, this is
another reason why the supergravity limit of the AdS/CFT correspondence
 works only in the limit of 
large $N$ \cite{o}.

It is worthwhile to mention that, for finite $N$, the number of single-trace
chiral primary $n=4$ superfields is in one to one correspondence
with the odd de Rham cohomology classes $H^\ell$ ($3\leq \ell \leq 2N-1 $)
of the group manifold $SU(N)$.\footnote{We would like to thank Raymond Stora
 for a discussion on this point.}

If this would be the case also for the multitrace operators, then there would
 be only a finite number of additional short multiplets coming from the 
cohomology classes $H^\ell$ with $2N-1 < \ell \leq N^2 -1$.

Incidentally we remark that $N^2-1$  is essentially the central charge
of the $N=4$ superconformal algebra \cite{afgj,a} so the bound would be
similar to the case of two dimensional superconformal field theories 
\cite{ms}.

The window of ``multiparticle states'' chiral primaries would then be 
$\Delta s =N(N-2)$ and it would grow, for large $N$, as $N^2$, faster than
single particle states which grow like $N$.

For finite $N$, the fact that the number of ``single trace chiral primaries''
 is finite may be related to a stringy effect that is not seen in the
supergravity approximation.
It is analogous to the ``stringy exclusion principle''
 discussed by Maldacena and Strominger \cite{ms}.
\section*{Acknowledgements}
We would like to thank D. Anselmi, L. Girardello, M. Porrati, R. Stora
 and A. Zaffaroni for enlightening conversations.


\appendix
\section{Field content of the $n=4$ Konishi-multiplet}
\begin{table}[h]
  \begin{center}
    \leavevmode
\caption{$(0,0)$ fields:  }
\label{0,0}
    \begin{tabular}{|c|l|}
\hline
 $SU(4)$ & $(\theta$-component$)_{E_0}$\\
\hline 
1&$(\theta^0 \bar\theta^0)_\ell$; $(\theta^2 \bar\theta^2)_{\ell +2}$;
$(\theta^8 )_{\ell +4}$; $(\theta^4 \bar\theta^4)_{\ell +4}$;
$(\bar\theta^8 )_{\ell +4}$;
$(\theta^6 \bar\theta^6)_{\ell +6}$;
$(\theta^8 \bar\theta^8)_{\ell +8}$\\
10 & $(\theta^2 )_{\ell +1}$; $(\theta^2\bar\theta^4 )_{\ell +3}$;
$(\bar\theta^6)_{\ell +3}$; $(\theta^4\bar\theta^6 )_{\ell +5}$;
 $(\theta^2\bar\theta^8)_{\ell +5}$; $(\theta^8\bar\theta^6)_{\ell +7}$\\
$\bar{10}$ & $(\bar\theta^2 )_{\ell +1}$; $(\theta^4\bar\theta^2 )_{\ell +3}$;
$(\theta^6)_{\ell +3}$; $(\theta^6\bar\theta^4 )_{\ell +5}$;
 $(\theta^8\bar\theta^2)_{\ell +5}$; $(\theta^6\bar\theta^8)_{\ell +7}$\\
$20_R$ & $(\theta^4)_{\ell +2}$; $(\bar\theta^4)_{\ell +2}$;
$(\theta^6 \bar\theta^2)_{\ell +4}$; $(\theta^4 \bar\theta^4)_{\ell +4}$; 
$(\theta^2 \bar\theta^6)_{\ell +4}$; 
$(\theta^4\bar\theta^8)_{\ell +6}$; $(\theta^8\bar\theta^4)_{\ell +6}$\\
15 & $(\theta^2 \bar\theta^2)_{\ell +2}$; $(\theta^4 \bar\theta^4)_{\ell +4}$;
$(\theta^6 \bar\theta^6)_{\ell +6}$\\
84 & $(\theta^2 \bar\theta^2)_{\ell +2}$; $(\theta^4 \bar\theta^4)_{\ell +4}$;
$(\theta^6 \bar\theta^6)_{\ell +6}$\\
64 & $(\theta^4\bar\theta^2 )_{\ell +3}$; $(\theta^2\bar\theta^4 )_{\ell +3}$;
$(\theta^6\bar\theta^4)_{\ell +5}$; $(\theta^4\bar\theta^6)_{\ell +5}$\\
126 & $(\theta^2\bar\theta^4 )_{\ell +3}$; $(\theta^4\bar\theta^6)_{\ell +5}$\\
$\bar{126}$ & 
$(\theta^4\bar\theta^2)_{\ell +3}$; $(\theta^6\bar\theta^4)_{\ell +5}$\\
35 & $(\theta^2\bar\theta^6)_{\ell +4}$\\
$\bar{35}$ & $(\theta^6\bar\theta^2)_{\ell +4}$\\
45 & $(\theta^2\bar\theta^6)_{\ell +4}$\\
$\bar{45}$ & $(\theta^6\bar\theta^2)_{\ell +4}$\\
105 &  $(\theta^4 \bar\theta^4)_{\ell +4}$\\
175 &  $(\theta^4 \bar\theta^4)_{\ell +4}$\\
\hline
    \end{tabular}   
  \end{center}
\end{table}
\begin{table}[h]
  \begin{center}
\caption{$(1,0)$ fields:  }
\label{1,0}
    \begin{tabular}{|c|l|}
\hline
 $SU(4)$ & $(\theta$-component$)_{E_0}$\\
\hline 
6 & $(\theta^2)_{\ell +1}$; $(\theta^6)_{\ell +3}$;
$(\theta^4 \bar\theta^2)_{\ell +3}$; $(\theta^2 \bar\theta^4)_{\ell +3}$;
 $(\theta^6\bar\theta^8)_{\ell +5}$; $(\theta^6 \bar\theta^4)_{\ell +5}$;
$(\theta^4 \bar\theta^6)_{\ell +5}$;\\
& $(\theta^2\bar\theta^8)_{\ell +7}$\\
15 & $(\theta^4)_{\ell +2}$; $(\theta^2\bar\theta^2)_{\ell +2}$;
 $(\theta^6\bar\theta^2)_{\ell +4}$; $(\theta^4 \bar\theta^4)_{\ell +4}$;
$(\theta^2\bar\theta^6)_{\ell +4}$; $(\theta^6\bar\theta^6)_{\ell +6}$;
$(\theta^4 \bar\theta^8)_{\ell +6}$\\
$\bar{45}$ & $(\theta^2\bar\theta^2)_{\ell +2}$; $(\theta^6\bar\theta^2)_{\ell +4}$;
$(\theta^4\bar\theta^4)_{\ell +4}$\\
45 & $(\theta^4\bar\theta^4)_{\ell +4}$; $(\theta^2\bar\theta^6)_{\ell +4}$;
$(\theta^6\bar\theta^6)_{\ell +6}$\\
$\bar{10}$ & $(\theta^4 \bar\theta^2)_{\ell +3}$\\
10 & $(\theta^4 \bar\theta^6)_{\ell +5}$\\
64 &  $(\theta^4 \bar\theta^2)_{\ell +3}$; $(\theta^2 \bar\theta^4)_{\ell +3}$;
$(\theta^6 \bar\theta^4)_{\ell +5}$; $(\theta^4 \bar\theta^6)_{\ell +5}$\\
$\bar{70}$ & $(\theta^4 \bar\theta^2)_{\ell +3}$\\
70 & $(\theta^4 \bar\theta^6)_{\ell +5}$\\
50 & $(\theta^2 \bar\theta^4)_{\ell +3}$;
$(\theta^6 \bar\theta^4)_{\ell +5}$\\
$20_R$ & $(\theta^4 \bar\theta^4)_{\ell +4}$\\
175 & $(\theta^4 \bar\theta^4)_{\ell +4}$\\
\hline
    \end{tabular}    
\caption{$(\half,\half)$ fields:  }
\label{1/2,1/2}
    \begin{tabular}{|c|l|}
\hline
 $SU(4)$ & $(\theta$-component$)_{E_0}$\\
\hline 
1 &
 $(\theta \bar\theta )_{\ell +1}$; $(\theta^3 \bar\theta^3)_{\ell +3}$;
$(\theta^5 \bar\theta^5)_{\ell +5}$; $(\theta^7 \bar\theta^7)_{\ell +7}$\\
15 &
  $(\theta \bar\theta )_{\ell +1}$; $(\theta^5 \bar\theta )_{\ell +3}$;
 $(\theta^3 \bar\theta^3)_{\ell +3}$; $(\theta^3 \bar\theta^3)_{\ell +3}$;
 $(\theta \bar\theta^5 )_{\ell +3}$;\\
 & $(\theta^7 \bar\theta^3)_{\ell +5}$;
$(\theta^5 \bar\theta^5)_{\ell +5}$; $(\theta^5 \bar\theta^5)_{\ell +5}$;
$(\theta^3 \bar\theta^7)_{\ell +5}$; $(\theta^7 \bar\theta^7)_{\ell +7}$\\
6 &
 $(\theta^3 \bar\theta)_{\ell +2}$; $(\theta \bar\theta^3)_{\ell +2}$;
$(\theta^7 \bar\theta)_{\ell +4}$; $(\theta^5 \bar\theta^3)_{\ell +4}$;
$(\theta^3 \bar\theta^5)_{\ell +4}$; $(\theta \bar\theta^7)_{\ell +4}$;\\
& $(\theta^7 \bar\theta^5)_{\ell +6}$; $(\theta^5 \bar\theta^7)_{\ell +6}$\\
10 &
 $(\theta^3 \bar\theta)_{\ell +2}$; $(\theta^5 \bar\theta^3)_{\ell +4}$;
$(\theta^3 \bar\theta^5)_{\ell +4}$; $(\theta \bar\theta^7)_{\ell +4}$;
$(\theta^7 \bar\theta^5)_{\ell +6}$; \\
$\bar{10}$ &
$(\theta \bar\theta^3)_{\ell +2}$; $(\theta^7 \bar\theta)_{\ell +4}$; 
$(\theta^5 \bar\theta^3)_{\ell +4}$; $(\theta^3 \bar\theta^5)_{\ell +4}$;
$(\theta^5 \bar\theta^7)_{\ell +6}$\\
64 &
 $(\theta^3 \bar\theta)_{\ell +2}$; $(\theta \bar\theta^3)_{\ell +2}$;
$(\theta^5 \bar\theta^3)_{\ell +4}$; $(\theta^5 \bar\theta^3)_{\ell +4}$;
$(\theta^3 \bar\theta^5)_{\ell +4}$; $(\theta^3 \bar\theta^5)_{\ell +4}$;\\
 & $(\theta^7 \bar\theta^5)_{\ell +6}$; $(\theta^5 \bar\theta^7)_{\ell +6}$\\
$20_R$ & 
$(\theta^5 \bar\theta )_{\ell +3}$; $(\theta^3 \bar\theta^3)_{\ell +3}$;
 $(\theta \bar\theta^5 )_{\ell +3}$; $(\theta^7 \bar\theta^3)_{\ell +5}$;
$(\theta^5 \bar\theta^5)_{\ell +5}$; $(\theta^3 \bar\theta^7)_{\ell +5}$\\
45 &
$(\theta^3 \bar\theta^3)_{\ell +3}$; $(\theta \bar\theta^5 )_{\ell +3}$;
$(\theta^5 \bar\theta^5)_{\ell +5}$; $(\theta^3 \bar\theta^7)_{\ell +5}$\\
$\bar{45}$ & 
$(\theta^5 \bar\theta )_{\ell +3}$; $(\theta^3 \bar\theta^3)_{\ell +3}$;
 $(\theta^7 \bar\theta^3)_{\ell +5}$; $(\theta^5 \bar\theta^5)_{\ell +5}$\\
84 &
$(\theta^3 \bar\theta^3)_{\ell +3}$; $(\theta^5 \bar\theta^5)_{\ell +5}$\\
175 &
 $(\theta^3 \bar\theta^3)_{\ell +3}$; $(\theta^5 \bar\theta^5)_{\ell +5}$\\
50 &
$(\theta^5 \bar\theta^3)_{\ell +4}$; $(\theta^3 \bar\theta^5)_{\ell +4}$\\
70 &
$(\theta^5 \bar\theta^3)_{\ell +4}$\\
$\bar{70}$ &
 $(\theta^3 \bar\theta^5)_{\ell +4}$\\
126 &
 $(\theta^3 \bar\theta^5)_{\ell +4}$\\ 
$\bar{126}$ &
$(\theta^5 \bar\theta^3)_{\ell +4}$\\
\hline
    \end{tabular}    
  \end{center}
\end{table}
\begin{table}[h]
  \begin{center}
\caption{$(2,0)$ fields:  }
\label{2,0}
    \begin{tabular}{|c|l|}
\hline
 $SU(4)$ & $(\theta$-component$)_{E_0}$\\
\hline 
1 &
 $(\theta^4 )_{\ell +2}$\\
$\bar{10}$ &
$(\theta^4 \bar\theta^2)_{\ell +3}$\\
10 &
$(\theta^4 \bar\theta^6)_{\ell +5}$\\
 $20_R$ & 
$(\theta^4 \bar\theta^4)_{\ell +4}$\\
\hline
    \end{tabular}
\caption{$(\frac{3}{2},\half)$ fields:  }
\label{3/2,1/2}
    \begin{tabular}{|c|l|}
\hline
 $SU(4)$ & $(\theta$-component$)_{E_0}$\\
\hline 
6 &
 $(\theta^3 \bar\theta)_{\ell +2}$;  $(\theta^5 \bar\theta^3)_{\ell +4}$;
$(\theta^3 \bar\theta^5)_{\ell +4}$; $(\theta^5 \bar\theta^7)_{\ell +6}$\\
$bar{10}$ &
 $(\theta^3 \bar\theta)_{\ell +2}$; $(\theta^5 \bar\theta^3)_{\ell +4}$\\
10 &
 $(\theta^3 \bar\theta^5)_{\ell +4}$; $(\theta^5 \bar\theta^7)_{\ell +6}$\\
1 &
$(\theta^5 \bar\theta)_{\ell +3}$; $(\theta^3\bar\theta^7)_{\ell +5}$\\
15 &
  $(\theta^5 \bar\theta )_{\ell +3}$; $(\theta^3 \bar\theta^3)_{\ell +3}$; 
 $(\theta^5 \bar\theta^5)_{\ell +5}$; $(\theta^3 \bar\theta^7)_{\ell +5}$\\
$20_R$ & 
$(\theta^3 \bar\theta^3)_{\ell +3}$;  $(\theta^5 \bar\theta^5)_{\ell +5}$\\
$\bar{45}$ &
$(\theta^3 \bar\theta^3)_{\ell +3}$\\
45 &
$(\theta^5 \bar\theta^5)_{\ell +5}$\\
64 & $(\theta^5 \bar\theta^3)_{\ell +4}$; $(\theta^3 \bar\theta^5)_{\ell +4}$\\
\hline
    \end{tabular}    
\caption{$(1,1)$ fields:  }
\label{1,1}
    \begin{tabular}{|c|l|}
\hline
 $SU(4)$ & $(\theta$-component$)_{E_0}$\\
\hline 
1 & 
$(\theta^2 \bar\theta^2)_{\ell +2}$; $(\theta^6 \bar\theta^2)_{\ell +4}$;
$(\theta^4 \bar\theta^4)_{\ell +4}$; $(\theta^2 \bar\theta^6)_{\ell +4}$;
$(\theta^6 \bar\theta^6)_{\ell +6}$\\
15 &
$(\theta^2 \bar\theta^2)_{\ell +2}$; $(\theta^6 \bar\theta^2)_{\ell +4}$;
$(\theta^4 \bar\theta^4)_{\ell +4}$; $(\theta^4 \bar\theta^4)_{\ell +4}$;
$(\theta^2 \bar\theta^6)_{\ell +4}$; $(\theta^6 \bar\theta^6)_{\ell +6}$\\
$20_R$ &
$(\theta^2 \bar\theta^2)_{\ell +2}$; $(\theta^6 \bar\theta^2)_{\ell +4}$;
$(\theta^4 \bar\theta^4)_{\ell +4}$; $(\theta^2 \bar\theta^6)_{\ell +4}$;
$(\theta^6 \bar\theta^6)_{\ell +6}$\\
6 & 
$(\theta^4 \bar\theta^2)_{\ell +3}$; $(\theta^2 \bar\theta^4)_{\ell +3}$;
$(\theta^6 \bar\theta^4)_{\ell +5}$; $(\theta^4 \bar\theta^6)_{\ell +5}$\\
10 &
$(\theta^4 \bar\theta^2)_{\ell +3}$; $(\theta^2 \bar\theta^4)_{\ell +3}$;
$(\theta^6 \bar\theta^4)_{\ell +5}$; $(\theta^4 \bar\theta^6)_{\ell +5}$\\
$\bar{10}$ &
$(\theta^4 \bar\theta^2)_{\ell +3}$; $(\theta^2 \bar\theta^4)_{\ell +3}$;
$(\theta^6 \bar\theta^4)_{\ell +5}$; $(\theta^4 \bar\theta^6)_{\ell +5}$\\
64 &
$(\theta^4 \bar\theta^2)_{\ell +3}$; $(\theta^2 \bar\theta^4)_{\ell +3}$;
$(\theta^6 \bar\theta^4)_{\ell +5}$; $(\theta^4 \bar\theta^6)_{\ell +5}$\\
45 &
$(\theta^4 \bar\theta^4)_{\ell +4}$\\
$\bar{45}$ &
$(\theta^4 \bar\theta^4)_{\ell +4}$\\
84 &
$(\theta^4 \bar\theta^4)_{\ell +4}$\\
\hline
    \end{tabular}
    \end{center}
\end{table}
\begin{table}[h]
 \begin{center}
\caption{$(2,1)$ fields:  }
\label{2,1}
    \begin{tabular}{|c|l|}
\hline
 $SU(4)$ & $(\theta$-component$)_{E_0}$\\
\hline 
6 & 
$(\theta^4 \bar\theta^2)_{\ell +3}$; $(\theta^4 \bar\theta^6)_{\ell +5}$\\
15 &
$(\theta^4 \bar\theta^4)_{\ell +4}$\\
\hline
    \end{tabular}
\caption{$(\frac{3}{2},\frac{3}{2})$ fields:  }
\label{3/2,3/2}
    \begin{tabular}{|c|l|}
\hline
 $SU(4)$ & $(\theta$-component$)_{E_0}$\\
\hline
1 &
$(\theta^3 \bar\theta^3)_{\ell +3}$; $(\theta^5 \bar\theta^5)_{\ell +5}$\\
15 & 
$(\theta^3 \bar\theta^3)_{\ell +3}$; $(\theta^5 \bar\theta^5)_{\ell +5}$\\
6 &
$(\theta^5 \bar\theta^3)_{\ell +4}$; $(\theta^3 \bar\theta^5)_{\ell +4}$\\
10 &
$(\theta^5 \bar\theta^3)_{\ell +4}$\\
$\bar{10}$ &
$(\theta^3 \bar\theta^5)_{\ell +4}$\\
\hline
    \end{tabular}
\caption{$(2,2)$ fields:  }
\label{2,2}
    \begin{tabular}{|c|l|}
\hline
 $SU(4)$ & $(\theta$-component$)_{E_0}$\\
\hline
1 &
$(\theta^4 \bar\theta^4)_{\ell +4}$\\
\hline
    \end{tabular}
      \end{center}
\end{table}


\begin{table}[h]
  \begin{center}
    \leavevmode
\caption{$(\half,0)$ fields:  }
\label{1/2,0}
    \begin{tabular}{|c|l|}
\hline
 $SU(4)$ & $(\theta$-component$)_{E_0}$\\
\hline
4 &
$(\theta )_{\ell +\half}$; $(\theta^3 \bar\theta^2)_{\ell +\frac{5}{2}}$;
$(\theta^5 \bar\theta^4)_{\ell +\frac{9}{2}}$;
$(\theta \bar\theta^8)_{\ell +\frac{9}{2}}$;
$(\theta^7 \bar\theta^6)_{\ell +\frac{13}{2}}$\\
$\bar{4}$ &
$(\theta \bar\theta^2)_{\ell +\frac{3}{2}}$;
$(\theta^7)_{\ell +\frac{7}{2}}$;
$(\theta^3 \bar\theta^4)_{\ell +\frac{7}{2}}$;
$(\theta^5 \bar\theta^6)_{\ell +\frac{11}{2}}$;
$(\theta^7 \bar\theta^8)_{\ell +\frac{15}{2}}$\\
20 &
$(\theta^3)_{\ell +\frac{3}{2}}$; 
$(\theta^5 \bar\theta^2)_{\ell +\frac{7}{2}}$;
$(\theta^3 \bar\theta^4)_{\ell +\frac{7}{2}}$;
$(\theta \bar\theta^6)_{\ell +\frac{7}{2}}$;
$(\theta^7 \bar\theta^4)_{\ell +\frac{11}{2}}$;
$(\theta^5 \bar\theta^6)_{\ell +\frac{11}{2}}$;
$(\theta^3 \bar\theta^8)_{\ell +\frac{11}{2}}$\\
$\bar{20}$ &
$(\theta^5 )_{\ell +\frac{5}{2}}$;
$(\theta^3 \bar\theta^2)_{\ell +\frac{5}{2}}$;
$(\theta \bar\theta^4)_{\ell +\frac{5}{2}}$;
$(\theta^5 \bar\theta^4)_{\ell +\frac{9}{2}}$;
$(\theta^7 \bar\theta^2)_{\ell +\frac{9}{2}}$;
$(\theta^3 \bar\theta^6)_{\ell +\frac{9}{2}}$;
$(\theta^5 \bar\theta^8)_{\ell +\frac{13}{2}}$\\
$\bar{36}$ &
$(\theta\bar\theta^2)_{\ell +\frac{3}{2}}$;
$(\theta^5 \bar\theta^2)_{\ell +\frac{7}{2}}$;
$(\theta^3 \bar\theta^4)_{\ell +\frac{7}{2}}$;
$(\theta^5 \bar\theta^6)_{\ell +\frac{11}{2}}$\\
36 &
$(\theta^3 \bar\theta^2)_{\ell +\frac{5}{2}}$;
$(\theta^5 \bar\theta^4)_{\ell +\frac{9}{2}}$;
$(\theta^3 \bar\theta^6)_{\ell +\frac{9}{2}}$;
$(\theta^7 \bar\theta^6)_{\ell +\frac{13}{2}}$\\
140 &
$(\theta^3 \bar\theta^2)_{\ell +\frac{5}{2}}$;
$(\theta^5 \bar\theta^4)_{\ell +\frac{9}{2}}$\\
$\bar{140}$ &
$(\theta^3 \bar\theta^4)_{\ell +\frac{7}{2}}$;
$(\theta^5 \bar\theta^6)_{\ell +\frac{11}{2}}$\\
60 &
$(\theta\bar\theta^4)_{\ell +\frac{5}{2}}$;
$(\theta^5 \bar\theta^4)_{\ell +\frac{9}{2}}$;
$(\theta^3 \bar\theta^6)_{\ell +\frac{9}{2}}$\\
$\bar{60}$ &
$(\theta^5 \bar\theta^2)_{\ell +\frac{7}{2}}$;
$(\theta^3 \bar\theta^4)_{\ell +\frac{7}{2}}$;
$(\theta^7 \bar\theta^4)_{\ell +\frac{7}{2}}$\\
$\bar{84'}$ &
$(\theta^5 \bar\theta^2)_{\ell +\frac{7}{2}}$\\
$84'$ &
$(\theta^3 \bar\theta^6)_{\ell +\frac{9}{2}}$\\
$\bar{140'}$ &
$(\theta^3 \bar\theta^4)_{\ell +\frac{7}{2}}$\\
$140'$ &
$(\theta^5 \bar\theta^4)_{\ell +\frac{9}{2}}$\\
$20''$ &
$(\theta \bar\theta^6)_{\ell +\frac{7}{2}}$\\
$\bar{20''}$ &
$(\theta^7 \bar\theta^2)_{\ell +\frac{9}{2}}$\\
\hline
    \end{tabular}
     \end{center}
\end{table}
\begin{table}[h]
 \begin{center}
    \leavevmode
\caption{$(\frac{3}{2},0)$ fields:  }
\label{3/2,0}
    \begin{tabular}{|c|l|}
\hline
 $SU(4)$ & $(\theta$-component$)_{E_0}$\\
\hline
$\bar{4}$ &
$(\theta^3)_{\ell +\frac{3}{2}}$;
$(\theta^5 \bar\theta^2)_{\ell +\frac{7}{2}}$;
$(\theta^3 \bar\theta^8)_{\ell +\frac{11}{2}}$\\
4 &
 $(\theta^5)_{\ell +\frac{5}{2}}$;
$(\theta^3 \bar\theta^6)_{\ell +\frac{9}{2}}$;
 $(\theta^5 \bar\theta^8)_{\ell +\frac{13}{2}}$\\
$\bar{20}$ &
 $(\theta^3 \bar\theta^2)_{\ell +\frac{5}{2}}$;
$(\theta^5 \bar\theta^4)_{\ell +\frac{9}{2}}$\\
20 &
$(\theta^3 \bar\theta^4)_{\ell +\frac{7}{2}}$;
$(\theta^5 \bar\theta^6)_{\ell +\frac{11}{2}}$\\
$\bar{20''}$ &
 $(\theta^3 \bar\theta^2)_{\ell +\frac{5}{2}}$\\
$20''$ &
$(\theta^5 \bar\theta^6)_{\ell +\frac{11}{2}}$\\
$\bar{36}$ & 
$(\theta^5 \bar\theta^2)_{\ell +\frac{7}{2}}$\\
36 &
$(\theta^3 \bar\theta^6)_{\ell +\frac{9}{2}}$\\
$\bar{60}$ & 
$(\theta^3 \bar\theta^4)_{\ell +\frac{7}{2}}$\\
60 &
$(\theta^5 \bar\theta^4)_{\ell +\frac{9}{2}}$\\
\hline
    \end{tabular}
\caption{$(1,\half)$ fields:  }
\label{1,1/2}
    \begin{tabular}{|c|l|}
\hline
 $SU(4)$ & $(\theta$-component$)_{E_0}$\\
\hline
4 &
$(\theta^2 \bar\theta)_{\ell +\frac{3}{2}}$;
$(\theta^6 \bar\theta)_{\ell +\frac{7}{2}}$;
$(\theta^4 \bar\theta^3)_{\ell +\frac{7}{2}}$;
$(\theta^2 \bar\theta^5)_{\ell +\frac{7}{2}}$;
$(\theta^6 \bar\theta^5)_{\ell +\frac{11}{2}}$;
$(\theta^4 \bar\theta^7)_{\ell +\frac{11}{2}}$\\
$\bar{4}$ & 
$(\theta^4 \bar\theta)_{\ell +\frac{5}{2}}$;
$(\theta^2 \bar\theta^3)_{\ell +\frac{5}{2}}$;
$(\theta^6 \bar\theta^3)_{\ell +\frac{9}{2}}$;
$(\theta^4 \bar\theta^5)_{\ell +\frac{9}{2}}$;
$(\theta^2 \bar\theta^7)_{\ell +\frac{9}{2}}$;
$(\theta^6 \bar\theta^7)_{\ell +\frac{13}{2}}$\\
$\bar{20}$ & 
$(\theta^2 \bar\theta)_{\ell +\frac{3}{2}}$;
$(\theta^6 \bar\theta)_{\ell +\frac{7}{2}}$;
$(\theta^4 \bar\theta^3)_{\ell +\frac{7}{2}}$;
$(\theta^4 \bar\theta^3)_{\ell +\frac{7}{2}}$;
$(\theta^2 \bar\theta^5)_{\ell +\frac{7}{2}}$;\\
& $(\theta^6 \bar\theta^5)_{\ell +\frac{11}{2}}$;
$(\theta^4 \bar\theta^7)_{\ell +\frac{11}{2}}$\\
20 &
$(\theta^4 \bar\theta)_{\ell +\frac{5}{2}}$;
$(\theta^2 \bar\theta^3)_{\ell +\frac{5}{2}}$;
$(\theta^6 \bar\theta^3)_{\ell +\frac{9}{2}}$;
$(\theta^4 \bar\theta^5)_{\ell +\frac{9}{2}}$;
$(\theta^4 \bar\theta^5)_{\ell +\frac{9}{2}}$;
$(\theta^2 \bar\theta^7)_{\ell +\frac{9}{2}}$;\\
& $(\theta^6 \bar\theta^7)_{\ell +\frac{13}{2}}$\\
$\bar{36}$ &
$(\theta^4 \bar\theta)_{\ell +\frac{5}{2}}$;
$(\theta^2 \bar\theta^3)_{\ell +\frac{5}{2}}$;
$(\theta^6 \bar\theta^3)_{\ell +\frac{9}{2}}$;
$(\theta^4 \bar\theta^5)_{\ell +\frac{9}{2}}$;\\
36 &
$(\theta^4 \bar\theta^3)_{\ell +\frac{7}{2}}$;
$(\theta^2 \bar\theta^5)_{\ell +\frac{7}{2}}$;
$(\theta^6 \bar\theta^5)_{\ell +\frac{11}{2}}$;
$(\theta^4 \bar\theta^7)_{\ell +\frac{11}{2}}$\\
$\bar{60}$ &
$(\theta^2 \bar\theta^3)_{\ell +\frac{5}{2}}$;
$(\theta^6 \bar\theta^3)_{\ell +\frac{9}{2}}$;
$(\theta^4 \bar\theta^5)_{\ell +\frac{9}{2}}$\\
60 &
$(\theta^4 \bar\theta^3)_{\ell +\frac{7}{2}}$;
$(\theta^2 \bar\theta^5)_{\ell +\frac{7}{2}}$;
$(\theta^6 \bar\theta^5)_{\ell +\frac{11}{2}}$\\
$\bar{20''}$ &
$(\theta^4 \bar\theta^3)_{\ell +\frac{7}{2}}$\\
$20''$ &
$(\theta^4 \bar\theta^5)_{\ell +\frac{9}{2}}$\\
140 &
$(\theta^4 \bar\theta^3)_{\ell +\frac{7}{2}}$\\
$\bar{140}$ &
$(\theta^4 \bar\theta^5)_{\ell +\frac{9}{2}}$\\
\hline
    \end{tabular}
     \end{center}
\end{table}
\begin{table}[h]
  \begin{center}
\caption{$(\frac{3}{2},1)$ fields:  }
\label{3/2,1}
    \begin{tabular}{|c|l|}
\hline
 $SU(4)$ & $(\theta$-component$)_{E_0}$\\
\hline
4 &
 $(\theta^3 \bar\theta^2)_{\ell +\frac{5}{2}}$;
$(\theta^5 \bar\theta^4)_{\ell +\frac{9}{2}}$;
$(\theta^3 \bar\theta^6)_{\ell +\frac{9}{2}}$\\
$\bar{4}$ &
$(\theta^5 \bar\theta^2)_{\ell +\frac{7}{2}}$;
$(\theta^3 \bar\theta^4)_{\ell +\frac{7}{2}}$;
$(\theta^5 \bar\theta^6)_{\ell +\frac{11}{2}}$\\
$\bar{20}$ &
 $(\theta^3 \bar\theta^2)_{\ell +\frac{5}{2}}$;
$(\theta^5 \bar\theta^4)_{\ell +\frac{9}{2}}$;
$(\theta^3 \bar\theta^6)_{\ell +\frac{9}{2}}$\\
20 &
$(\theta^5 \bar\theta^2)_{\ell +\frac{7}{2}}$;
$(\theta^3 \bar\theta^4)_{\ell +\frac{7}{2}}$;
$(\theta^5 \bar\theta^6)_{\ell +\frac{11}{2}}$\\
$\bar{36}$ &
$(\theta^3 \bar\theta^4)_{\ell +\frac{7}{2}}$\\
36 &
$(\theta^5 \bar\theta^4)_{\ell +\frac{9}{2}}$\\
\hline
    \end{tabular}
\caption{$(2,\half)$ fields:  }
\label{2,1/2}
    \begin{tabular}{|c|l|}
\hline
 $SU(4)$ & $(\theta$-component$)_{E_0}$\\
\hline
$\bar{4}$ & 
$(\theta^4 \bar\theta)_{\ell +\frac{5}{2}}$\\
4 & 
$(\theta^4 \bar\theta^7)_{\ell +\frac{11}{2}}$\\
$\bar{20}$ & 
$(\theta^4 \bar\theta^3)_{\ell +\frac{7}{2}}$\\
20 &
$(\theta^4 \bar\theta^5)_{\ell +\frac{9}{2}}$\\
\hline
    \end{tabular}
\caption{$(2,\frac{3}{2})$ fields:  }
\label{2,3/2}
    \begin{tabular}{|c|l|}
\hline
 $SU(4)$ & $(\theta$-component$)_{E_0}$\\
\hline
4 &
$(\theta^4 \bar\theta^3)_{\ell +\frac{7}{2}}$\\
$\bar{4}$ & 
$(\theta^4 \bar\theta^5)_{\ell +\frac{9}{2}}$\\
\hline
    \end{tabular}
     \end{center}
\end{table}



\begin{thebibliography}{50}
\bibitem{m}
J.M. Maldacena, hep-th/9711200; 
S.S. Gubser, I.R. Klebanov and A.M. Polyakov, Phys. Lett. {\bf B428} (1998)
105, hep-th/9802109; 
E. Witten,
Adv. Theor. Math. Phys. 2 (1998) 253, 
 hep-th/9802150. 
\bibitem{bg}
T. Banks and M.B. Green,
 J.High Energy Phys. 05 (1998) 002, hep-th/9804170;
M. Bianchi, M.B. Green, Stefano Kovacs and G. Rossi,
 J.High Energy Phys. 9808 (1998) 013, hep-th/9807033;
J.H. Brodie and M. Gutperle, hep-th/9809067. 
\bibitem{fz}
S. Ferrara, and A. Zaffaroni, hep-th/9807090;
 L. Andrianopoli and S. Ferrara, hep-th/9807150;
S. Ferrara, M.A. Lledo and A. Zaffaroni, Phys. Rev. {\bf D58} (1998) 10.
\bibitem{afgj}
D. Anselmi, D.Z. Freedman, M.T. Grisaru and A.A. Johansen,
Phys. Lett {\bf B394} (1997) 329, hep-th/9608125.
\bibitem{a}
D. Anselmi, hep-th/9808004; hep-th/9809192; hep-th/9811149.
\bibitem{ls}
R.G. Leigh and M.J. Strassler, Nucl. Phys. {\bf B447} (1995) 95, 
 hep-th/9503121.
\bibitem{kw}
 I.R. Klebanov and E. Witten, hep-th/9807080;
 S.S. Gubser, hep-th/9807164.
\bibitem{gppz}
 L. Girardello, M. Petrini, M. Porrati and A. Zaffaroni, hep-th/9810126.
\bibitem{dz}
J. Distler and F. Zamora, hep-th/9810206.
\bibitem{kpw}
A. Khavaev, K. Pilch and N.P. Warner, hep-th/9812035.
\bibitem{cfn}
A. Ceresole, P. Fr\'e and H. Nicolai, Class. Quant. Grav. 2 (1985) 133.
\bibitem{gs}
M. G\"unaydin and C. Saclioglu, Commun. Math. Phys. 87 (1982) 159.
\bibitem{gm}
 M. G\"unaydin and  N. Marcus, Class. Quant. Grav. 2 (1985) L11.
\bibitem{hw}
P.S. Howe and P.C. West, Phys. Lett. {\bf B389} (1996) 273;
 Phys. Lett. {\bf B400} (1997) 307.  
\bibitem{af}
L. Andrianopoli and S. Ferrara, Phys. Lett. {\bf B430} (1998) 248, 
 hep-th/9803171.
\bibitem{ff}
S. Ferrara and C. Fronsdal, Phys. Lett. {\bf B433} (1998) 19;
S. Ferrara, C. Fronsdal and A. Zaffaroni, Nucl. Phys. {\bf B532} (1998) 153,
 hep-th/9802203.
\bibitem{fgg}
S. Ferrara, R. Gatto and A.F. Grillo, Phys. Rev. {\bf D9} (1974) 3564. 
\bibitem{hst}
P. Howe, K.S. Stelle and P.K. Townsend, Nucl. Phys. {\bf B192} (1981) 332.
\bibitem{gmz}
M. Gunaydin, D. Minic and M. Zagermann, Nucl. Phys. {\bf B534} (1998) 96,
 hep-th/9806042.
\bibitem{o}
H. Osborn, hep-th/9808041;
H. Liu, hep-th/9811152;
 S. Lee, S. Minwalla, M. Rangamani and N. Seiberg, hep-th/9806074;
H. Liu and A.A. Tseytlin, hep-th/9807097;
D.Z. Freedman, S.D. Mathur, A Matusis and L. Rastelli, hep-th/9808006;
T. Banks, M.R. Douglas, G.T. Horowitz and E. Martinec, hep-th/9808016;
B. Eden, P.S. Howe, C. Schubert, E. Sokatchev and P.C. West, hep-th/9811172.
\bibitem{gikos}
A. Galperin, E. Ivanov, S. Kalitsyn, V. Ogievetskii, E. Sokatchev,
 Class. Quant. Grav. 1 (1984) 469       
\bibitem{ms}
J.M. Maldacena and A. Strominger, hep-th/9804085.

 
\end{thebibliography}
\end{document}